\documentclass[12pt]{article}
\usepackage{graphicx}
\usepackage{lineno}

\newcommand\pubnumber{---}
\newcommand\pubdate{\today}

\def\florence{Department of Physics\\
University of Florence, 
INFN of Florence, ITALY}

\def\Title#1{\begin{center} {\Large #1 } \end{center}}
\def\Author#1{\begin{center}{ \sc #1} \end{center}}
\def\Address#1{\begin{center}{ \it #1} \end{center}}

\newcommand\pubblock{\rightline{\begin{tabular}{l} \pubnumber\\
         \pubdate  \end{tabular}}}
\newenvironment{Abstract}{\begin{quotation}  }{\end{quotation}}
\newenvironment{Presented}{\begin{quotation} \begin{center} 
             PRESENTED AT\end{center}\bigskip 
      \begin{center}\begin{large}}{\end{large}\end{center} \end{quotation}}

\def\beq{\begin{equation}}
\def\eeq#1{\label{#1}\end{equation}}
\def\eeqn{\end{equation}}

\def\beqa{\begin{eqnarray}}
\def\eeqa#1{\label{#1}\end{eqnarray}}
\def\eeqan{\end{eqnarray}}

\let\bar=\overbar

\def\Dslash{\not{\hbox{\kern-4pt $D$}}}
\def\dslash{\not{\hbox{\kern-2pt $\del$}}}

\def\msb{{\bar{\ssstyle M \kern -1pt S}}}

\linenumbers

\begin{document}
\begin{titlepage}
\pubblock

\vfill
\Title{$J/\psi$ production and polarization}
\vfill
\Author{Maddalena Frosini \\ 
on behalf of LHCb Collaboration}
\Address{\florence}
\vfill
\begin{Abstract}
The study of the production of heavy quarkonium is crucial for a thorough understanding of Quantum Chromodynamics (QCD). This note reports the measurements of the $J/\psi$, $\chi _{c}$
and double charm production cross section, and discusses the prospects for the $J/\psi$ polarization at LHCb.
\end{Abstract}
\vfill
\begin{Presented}
CHARM 2012\\
Honolulu, Hawaii,  May 14--17, 2012
\end{Presented}
\vfill
\end{titlepage}
\def\thefootnote{\fnsymbol{footnote}}
\setcounter{footnote}{0}

\section{Introduction}

Understanding the charmonium hadroproduction mechanism has been a long-term program both experimentally and theoretically. At the LHC two different components contribute: the ``prompt" component, which includes ``direct" production in the $pp$ collisions and the ``feed down" from higher charmonium states, and the ``delayed" component, coming from the $b$-hadron decays. In the direct production the $c\overline{c}$ pairs are expected to be created mainly through gluon-gluon fusion at the leading order (LO) and the bound states are formed in the final color-singlet states. The most recent models allow the formation of the $c\overline{c}$ pairs also through color-octet states, which evolve toward the final state via exchange of soft gluons. Such evolution is described in terms of a non-perturbative QCD (NRQCD) factorization approach. \\
The LHCb collaboration has given many contributions in understanding the quarkonium production mechanism, some of them reported here. The measurement of the $J/\psi$ production cross section together with the status of the polarization analysis will be presented. Both measurements provide a critical test for the color-singlet \cite{ref:CSM1, ref:CSM2} and color-octet models \cite{ref:COM}. 
The study of $\chi _{c}$, double $J/\psi$ and $J/\psi$ production associated with an open charm hadron will also be presented. In particular the last two are rare processes and they can be useful to investigate the contributions from other mechanisms, such as the Double Parton Scattering (DPS) \cite{DPS1, DPS2, DPS3}.

\section{The LHCb detector}

The LHCb detector~\cite{LHCbDec} is a single-arm forward spectrometer 
covering the pseudo-rapidity range $2 < \eta < 5$, designed for the study of
particles containing $b$ or $c$ quarks.
The detector includes a high precision tracking system consisting of a
silicon-strip vertex detector surrounding the proton-proton interaction
region, a large-area silicon-strip detector located upstream of a dipole
magnet with a bending power of about 4~Tm, and three stations of silicon-strip 
detectors and straw drift-tubes placed downstream. 
The combined tracking system has a momentum resolution $\delta p/p$ that
varies from 0.4\% at 5 GeV/$c^{2}$ to 0.6\% at 100 GeV/$c^{2}$, and an impact parameter
resolution of 20~$\mu$m for tracks with high transverse momentum. 
Charged hadrons are identified using two ring-imaging Cherenkov detectors.
Photon, electron and hadron candidates are identified by a calorimeter system 
consisting of scintillating-pad and pre-shower detectors, an electromagnetic
calorimeter and a hadron calorimeter. 
Muons are identified by a muon system composed of alternating layers of iron
and multiwire proportional chambers. 
The trigger consists of a hardware stage, based on information from the
calorimeter and muon systems, followed by a software stage which applies a
full event reconstruction.

\section{$J/\psi$ cross section measurement}

The cross section is measured selecting $J/\psi$ decaying to two muons: the data sample corresponds to an integrated luminosity $\mathcal{L}= \left(  5.2 \pm 0.5 \right) $ pb$^{-1}$ of $pp$ collisions at $\sqrt{s} = 7$ TeV recorded by the experiment during September 2010. The double differential cross section in $J/\psi$ $p_{T}$ and $y$ is defined as following:
\begin{equation}
\frac{d^{2} \sigma}{dp_{T} dy} = \frac{N (J/\psi \rightarrow \mu ^{+} \mu ^{-})}{L \times \varepsilon _{tot} \times \mathcal{B}(J/\psi \rightarrow \mu ^{+} \mu ^{-}) \Delta p_{T} \Delta y} ,
\end{equation}
where $N (J/\psi \rightarrow \mu ^{+} \mu ^{-})$ is the number of selected $J/\psi$ decaying in two muons, $\mathcal{L}$ is the integrated luminosity, $\varepsilon _{tot}$ is the total efficiency (estimated from Monte Carlo including the detector acceptance, the reconstruction and trigger efficiency), $\mathcal{B}(J/\psi \rightarrow \mu ^{+} \mu ^{-})$ is the branching ratio of the $J/\psi \rightarrow \mu ^{+} \mu ^{-}$ decay, $\Delta p_{T}$ and $\Delta y$ are respectively the $J/\psi$ transverse momentum and rapidity bin sizes.
The analysis selection requires at least one reconstructed primary vertex in each event. The $J/\psi$ candidates are formed from pairs of opposite sign charged tracks reconstructed in the tracking system and identified as muons. The two muons must have a good quality of the track fit and originate from a common vertex. To separate the prompt and the delayed component the $J/\psi$ pseudo proper time is used, defined as $
t_{z} = \frac{(z_{J/\psi} - z_{PV}) m_{J/\psi}}{p_{z}} ,
$
where $z_{J/\psi}$ and $z_{PV}$ are the $J/\psi$ decay vertex and the primary vertex positions along the beam axis and $m_{J/\psi}$ and $p_{z}$ are respectively the mass and the momentum component of the $J/\psi$ along the beam axis.

\begin{figure}
\begin{minipage}{.48\textwidth}
\centering
\includegraphics[width=1.\textwidth]{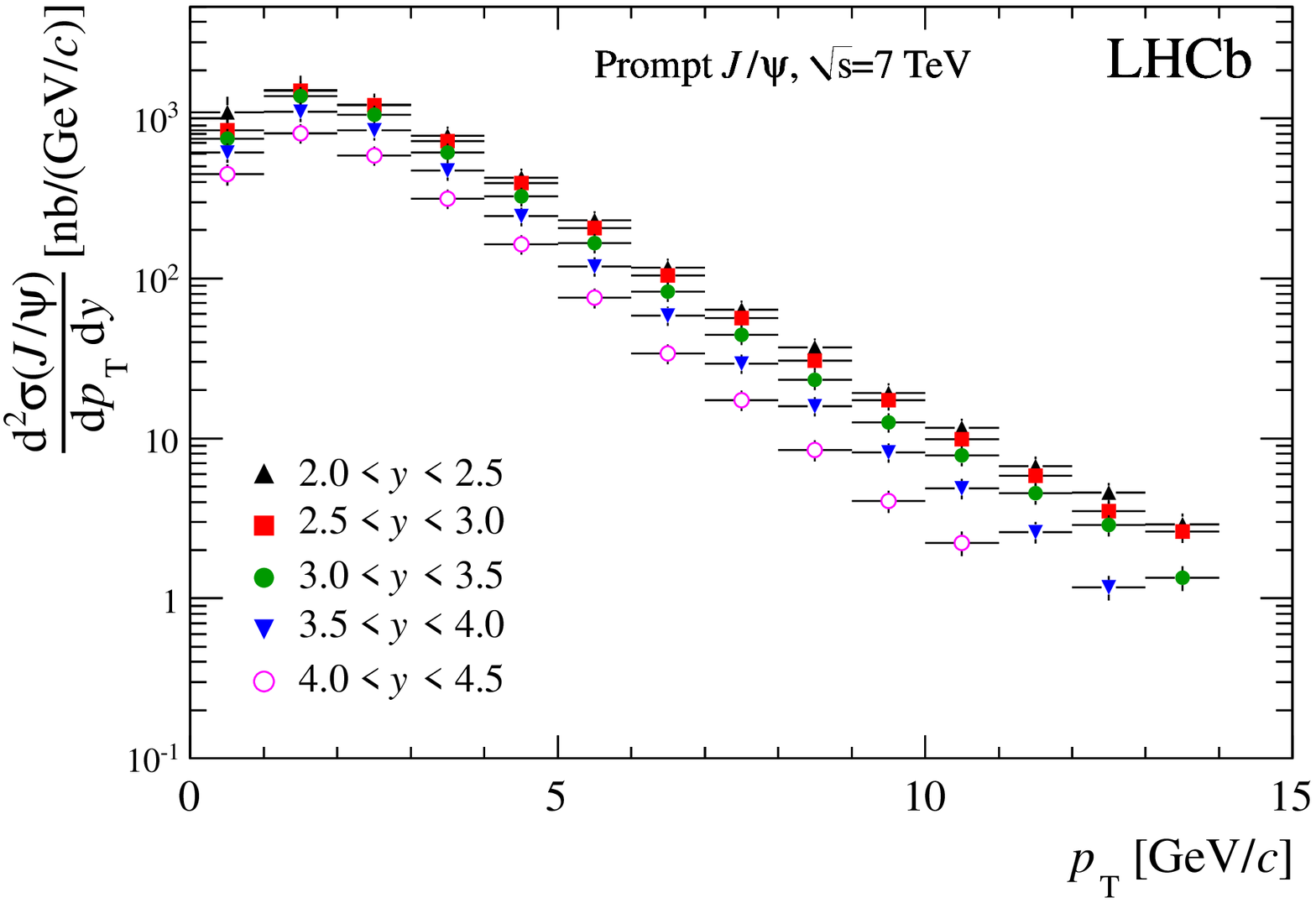}
\caption{Double differential cross section of $J/\psi$ prompt component.}\label{ddprompt}
\end{minipage}
\begin{minipage}{.49\textwidth}
\centering
\includegraphics[width=1.\textwidth]{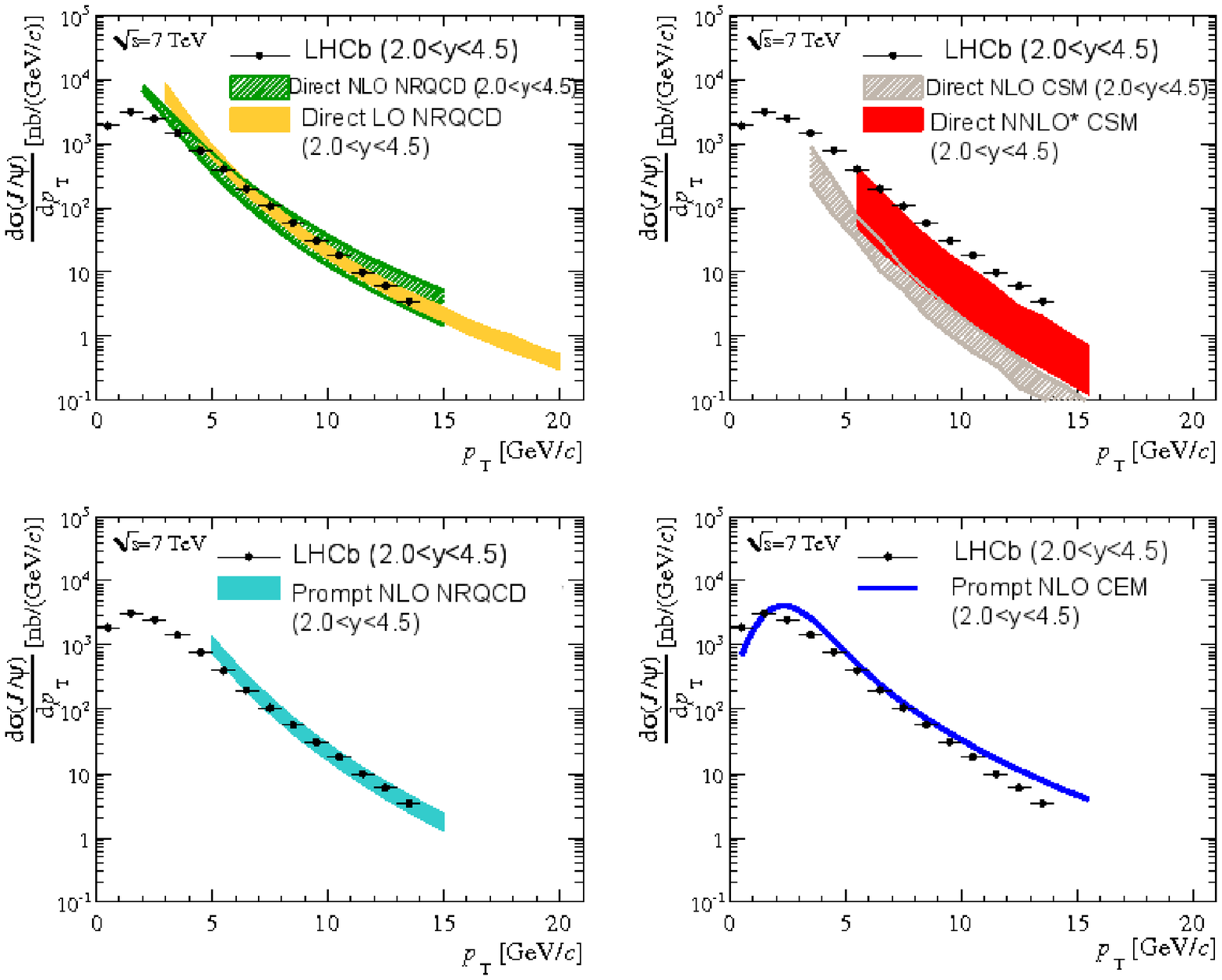}
\caption{Prompt $J/\psi$ differential cross section compared with different theoretical models. The top plots show the direct component only and the bottom include the feed down.}
\label{dprompttheory}
\end{minipage}
\end{figure}
\begin{figure}
\begin{minipage}{0.48\textwidth}
\centering
\includegraphics[width=1.\textwidth]{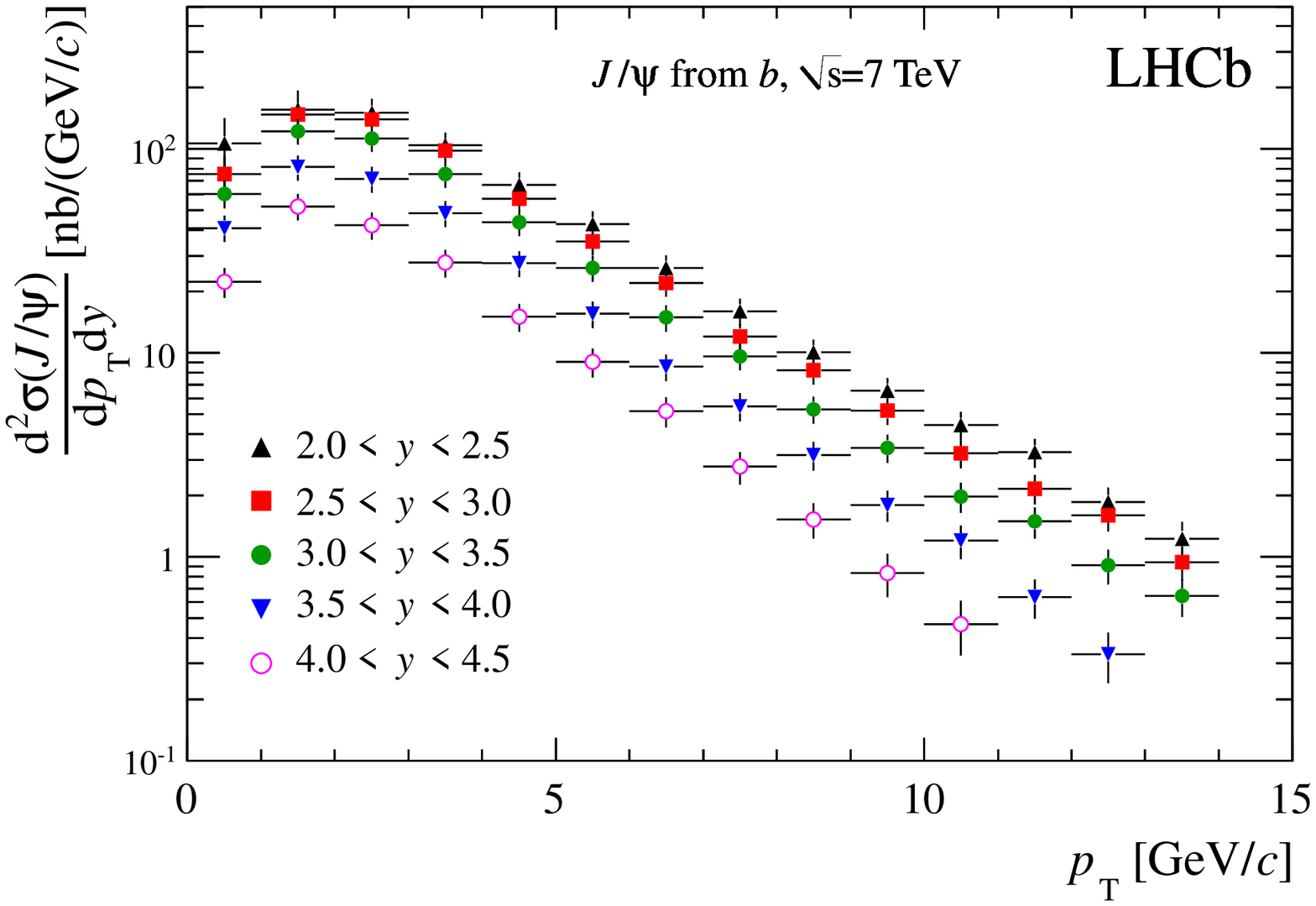}
\caption{Double differential cross section of the delayed $J/\psi$ component.}
\label{dddel}
\end{minipage}
\begin{minipage}{0.48\textwidth}
\includegraphics[width=1.\textwidth]{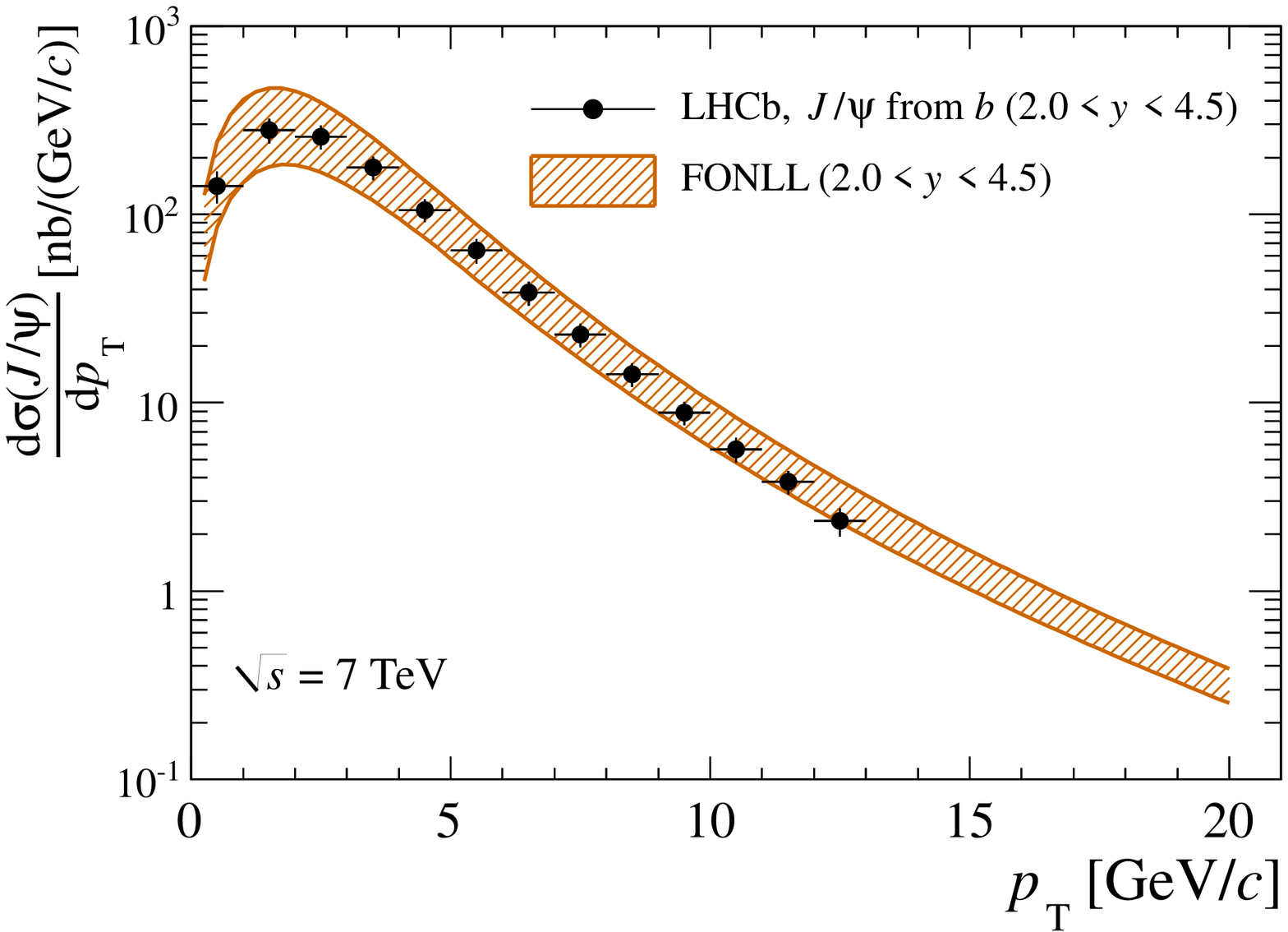}
\caption{Differential cross section of delayed $J/\psi$ component compared with FONLL computation.}
\label{ddeltheory}
\end{minipage}
\end{figure}
Figures \ref{ddprompt} and \ref{dprompttheory} show respectively the double differential prompt cross section and the differential prompt cross section integrated over rapidity as a function of $p_{T}$. Results are compared with the prediction of three different theoretical models (Colour Singlet Model, Colour Octet Model and Colour Evaporation Model). In Fig.\ref{dddel} the double differential cross section of the delayed component is shown and in Fig.\ref{ddeltheory} it is integrated over rapidity and compared with the FONLL computation. The total integrated cross sections are
\begin{equation}
\sigma _{prompt} = (10.52\pm 0.04 (stat.) \pm 1.40 (sys.) ^{+1.64} _{-2.20} (pol.)) \mu \mathrm{b} ,
\end{equation} 
\begin{equation}
\sigma _{from \: b} = \left[ 1.14\pm 0.01 (stat.) \pm 0.16 (sys.)\right]  \mu \mathrm{b} .
\label{totdel}
\end{equation}
The first and the second uncertainties are the statistical and the systematic, where the main sources of systematic uncertainty come from the luminosity measurement, the tracking and trigger efficiency. The third uncertainty on the prompt cross section is due to the unknown polarization of the $J/\psi$ and it is estimated calculating the total efficiency in two possible extreme scenarios of fully transverse and fully longitudinal polarization. The deviation from the case of zero polarization is assigned as systematic uncertainty to the measurement. From Eq.\ref{totdel} the $b\overline{b}$ cross section is extrapolated to the full solid angle using the formula
\begin{equation}
\sigma (pp \rightarrow b\overline{b} X) = \alpha _{4\pi} \frac{\sigma _{from b} } {2\mathcal{B} (b \rightarrow J/\psi X)} = \left[  288 \pm 4 (stat.)\pm 48 (sys.) \right]   \mu \mathrm{b} .
\end{equation}
All these results have been published in Ref.\cite{ref:pap}.

\section{Outlook for polarization measurement}

With the full 2011 data sample the prompt $J/\psi$ polarization will be measured studying the full angular distribution of the two muons:
\begin{equation}
\frac{dN}{d(\cos \theta) d\phi} \propto 1 + \lambda _{\theta} \cos ^{2} \theta + \lambda _{\phi} \sin ^{2} \theta \cos 2\phi +  \lambda _{\theta \phi} \sin 2\theta \cos \phi ,
\end{equation}
where $\theta$ and $\phi$ are the polar and azimuthal angles in the helicity frame (using the $J/\psi$ momentum as polarization axis). The measurement will be performed in bins of $J/\psi$ transverse momentum and rapidity. The statistical sensitivity should be under 0.15 for $\lambda _{\theta}$ and about 0.01 for $\lambda _{\phi}$ and $\lambda _{\theta \phi}$. The systematic uncertainty is expected to be of the same order of magnitude. 

\section{$\chi _{c}$ production}
The study of the $J/\psi$ production through the radiative decays of the $\chi _{c}$ states provides a useful test of both the color-singlet and color-octet model. Moreover it is fundamental for the $J/\psi$ polarization measurement, as the directly produced $J/\psi$ and those coming from $\chi _{c}$ decays can carry different polarization and this represents a possible source of uncertainty for the polarization measurement of the prompt component. The measurement of the fraction of $J/\psi$ coming from
$\chi _{c}$ decays can quantify this uncertainty.\\
\begin{figure}
\begin{minipage}{.48\textwidth}
\centering
\includegraphics[width=1.\textwidth]{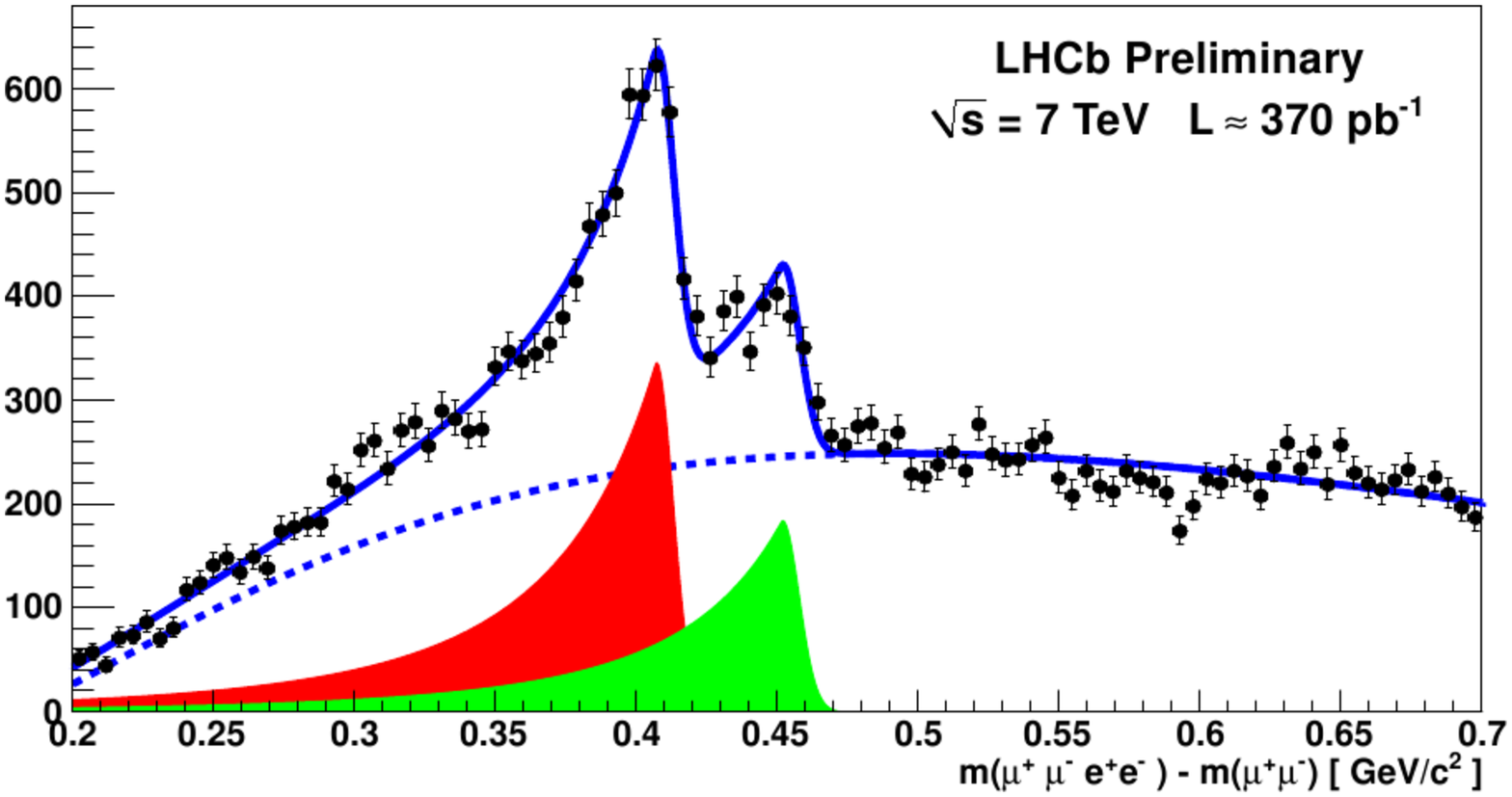}
\caption{Invariant mass difference spectrum $\Delta m = m(\chi _{c}) - m(J/\psi) = m(e^{+} e^{-} \mu ^{+} \mu^{-}) - m(\mu ^{+} \mu^{-})$ using converted photons.}\label{chicyield}
\end{minipage}
\begin{minipage}{.49\textwidth}
\centering
\includegraphics[width=1.\textwidth]{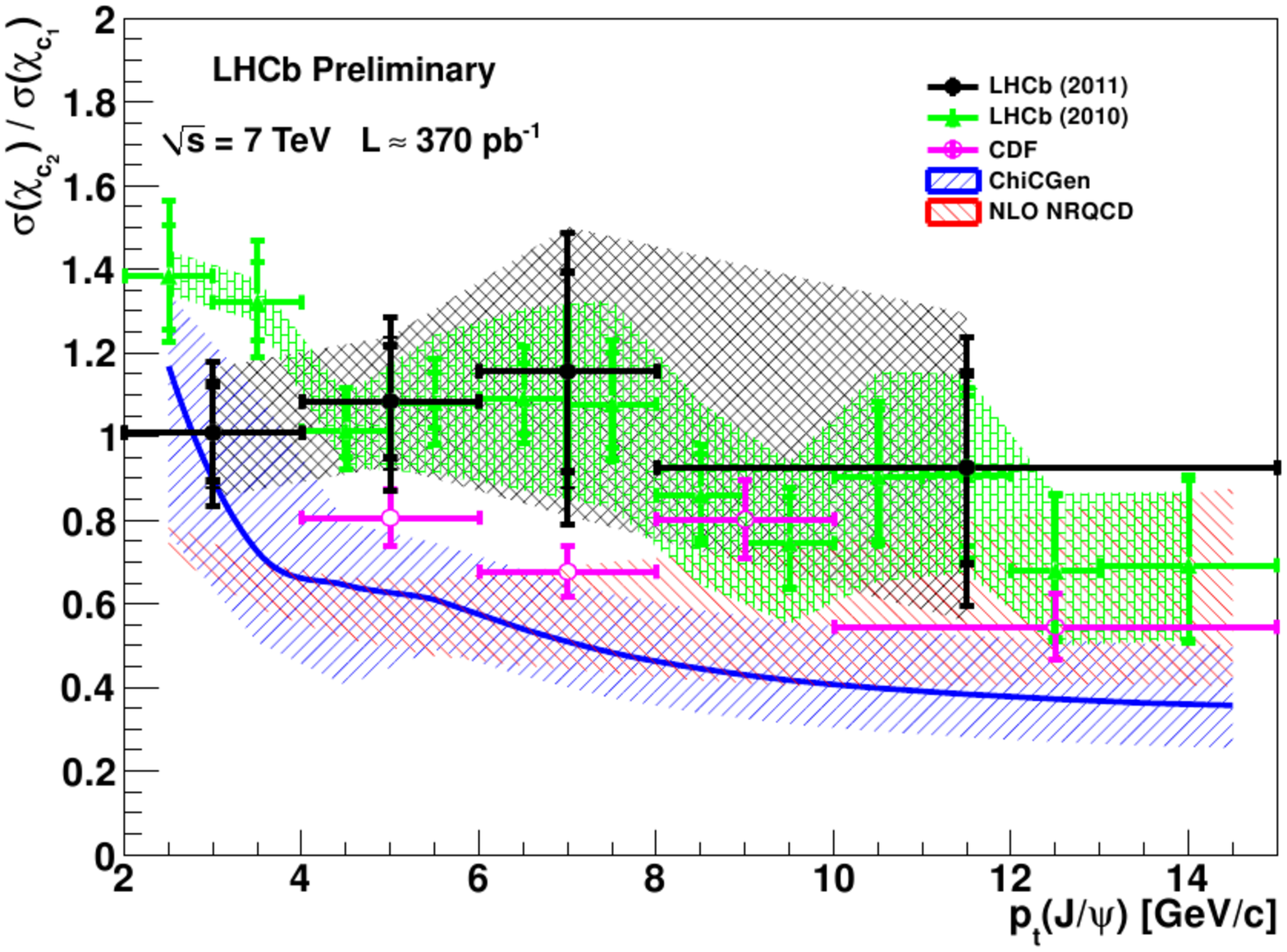}
\caption{Relative cross section $\sigma (\chi _{c2})/ \sigma (\chi _{c1})$ in bins of $J/\psi$ transverse momentum.}
\label{chicxsec}
\end{minipage}
\end{figure}
The relative cross section $\sigma (\chi _{c2})/ \sigma (\chi _{c1})$ is measured using two different data sample acquired by the LHCb experiment during the 2010 and 2011, respectively of 37 pb$^{-1}$ \cite{chic21noconv} and 370 pb$^{-1}$ \cite{chic21conv}. In both cases the $\chi _{c}$ states are identified through their radiative decay $\chi _{c} \rightarrow J/\psi \gamma$ with the $J/\psi$ decaying to two muons $J/\psi \rightarrow \mu ^{+} \mu ^{-}$. For the first measurement, made with a smaller data sample, the photons reconstructed in the calorimeter system have been used. This allows to have a higher statistics but the poor resolution of the calorimeter doesn't permit to separate the two $\chi _{c1}$ and $\chi _{c2}$ states. In the second measurement the photons converted in the detector material before the magnet have been used, $\gamma \rightarrow e^{+} e^{-}$. In this way it is possible to take advantage of the good resolution of the tracker, which allows to resolve the two states, as it is shown in Fig.\ref{chicyield}.\\
In both measurements the efficiency is determined from the Monte Carlo simulation and the number of signal events is extracted with a fit to the invariant mass difference spectra, in four bins of $J/\psi$ transverse momentum.\\
The results for the relative cross section $\sigma (\chi _{c2})/ \sigma (\chi _{c1})$ are shown in Fig.\ref{chicxsec}. In the plot the inner error bars correspond to the statistical uncertainties and the outer bars correspond to the sum of all the sources of systematic uncertainties. The shaded area represents the maximum effect due to the unknown $\chi _{c}$ polarization. In green and black the results obtained reconstructing the photons in the calorimeter (2010 statistics) and the converted photons (2011 statistics) are shown. The results from the CDF collaboration are shown in magenta \cite{CDFchic2}. The blue and red shaded area correspond to the color-singlet and NRQCD prediction respectively.

\begin{figure}
\begin{minipage}{.48\textwidth}
\centering
\includegraphics[width=1.\textwidth]{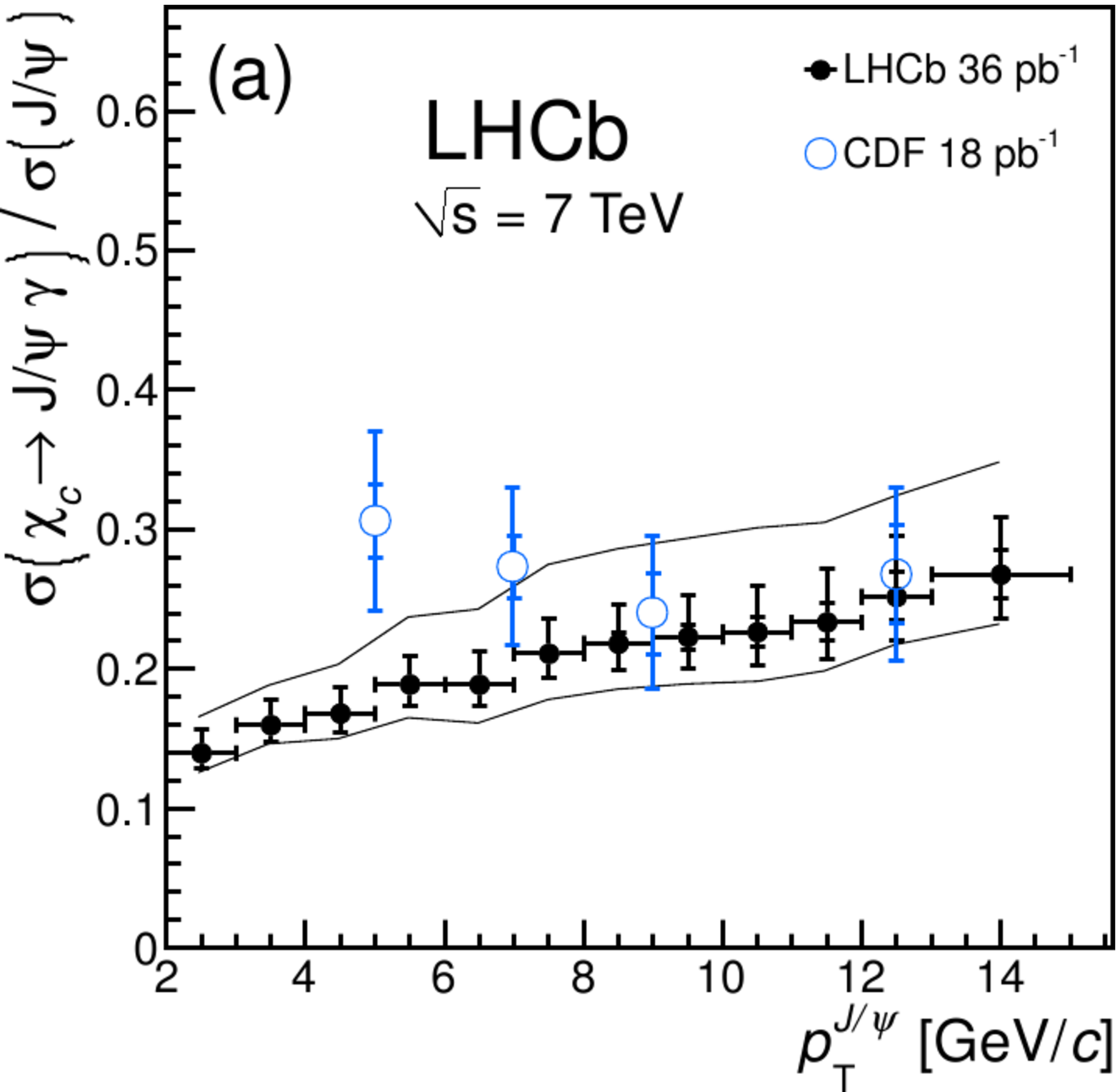}
\end{minipage}
\begin{minipage}{.48\textwidth}
\centering
\includegraphics[width=1.\textwidth]{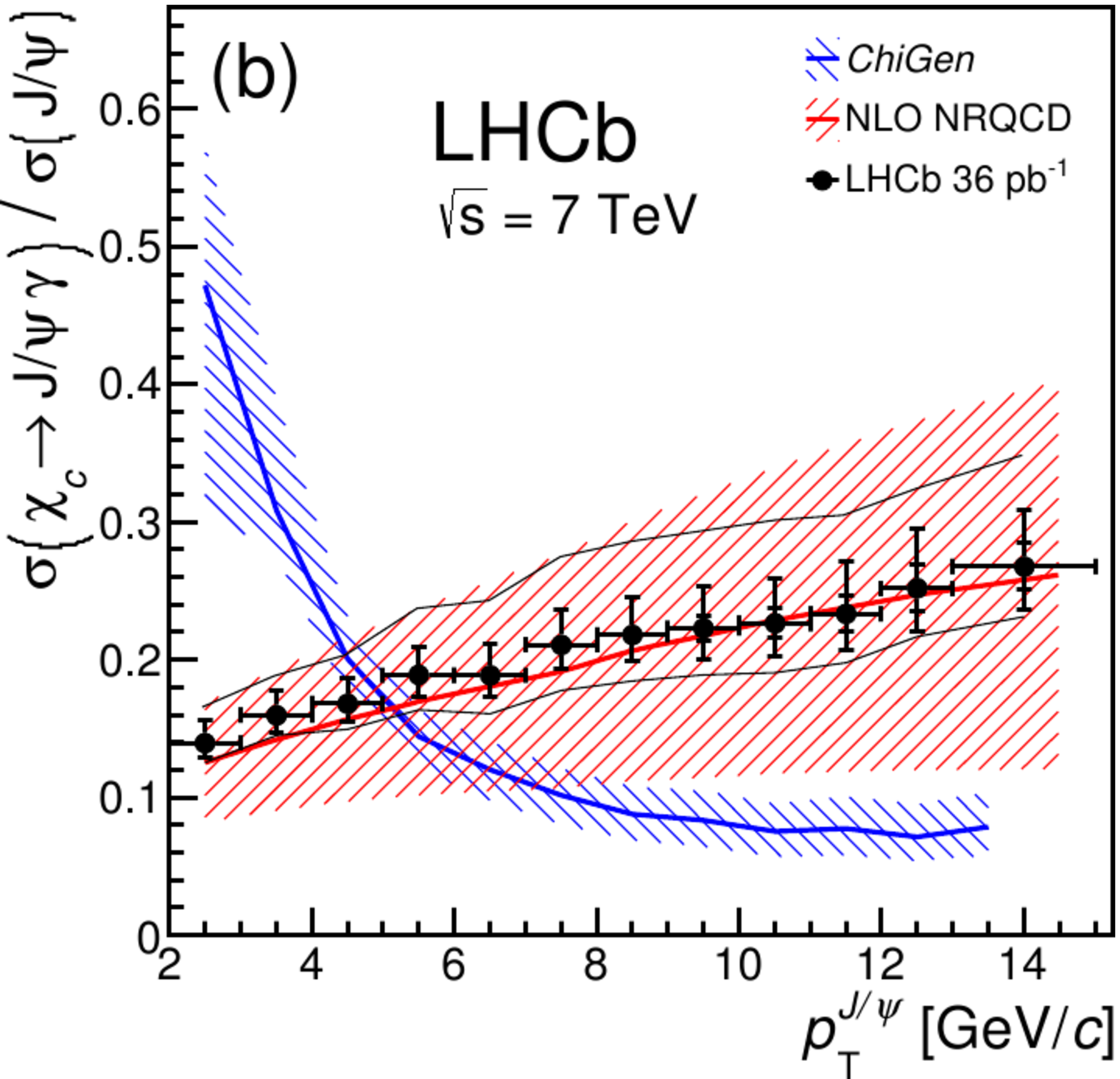}
\end{minipage}
\caption{Relative cross section $\sigma (\chi _{c})/ \sigma (J/\psi)$ in bins of $J/\psi$ transverse momentum. The results (black points) are compared with the CDF measurement \cite{CDFchic1} and with the color-singlet and color-octet prediction (respectively the blue and red area.)}\label{chicjpsi}
\end{figure}
The $\chi _{c}$ to $J/\psi$ ratio has been measured with the 36 pb$^{-1}$ data sample acquired by the experiment in the 2010 \cite{chicjpsi}. The $\chi _{c}$ states are reconstructed through their radiative decay $\chi _{c} \rightarrow J/\psi \gamma$ with the $J/\psi$ decaying into two muons $J/\psi \rightarrow \mu ^{+} \mu ^{-}$. The results are shown in Fig.\ref{chicjpsi}, compared with the CDF measurement and with two theoretical models, the color-singlet (blue) and NRQCD approach (red). 

\section{Double charm production}
Recently, both the production of double quarkonium and also the associated production of quarkonium together with open charm have been suggested as probes of the production mechanism. In the $pp$ collisions also other mechanisms, as the DPS \cite{DPS1, DPS2, DPS3}, can be involved in the production and their contribution can be estimated with respect to the Single Parton Scattering (SPS) \cite{SPS1, SPS2, SPS3}.\\
Such a study has been performed at LHCb through the measurement of the double $J/\psi$, taken from Ref.\cite{2jpsi}, and $J/\psi$ production associated with an open charm hadron (such as $D^{0}$, $D^{+}$, $D^{+}_{s}$ and $\Lambda _{c}^{+}$) \cite{2jpsiopen}, with the 2010 and 2011 datasets (respectively 37 pb$^{-1}$ and 355 pb$^{-1}$).

\begin{figure}
\centering
\includegraphics[width=.6\textwidth]{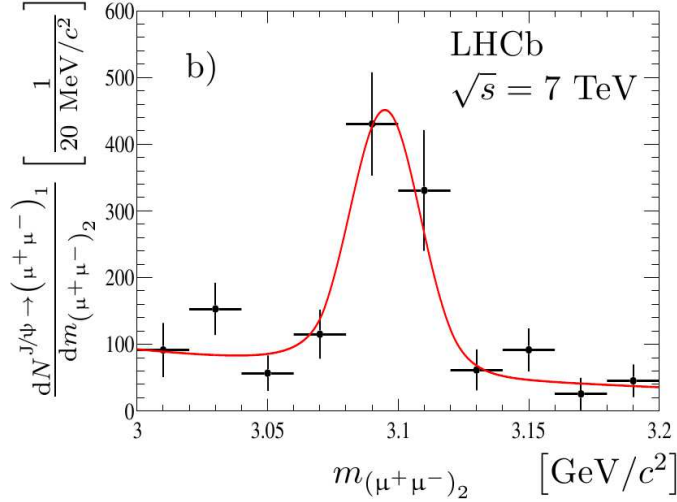}
\caption{Invariant mass distribution of the first muon pair in bins of the second muon pair for the double $J/\psi$ production.}
\label{2jpsiyield}
\end{figure}
The double $J/\psi$ production cross section has been measured reconstructing the two $J/\psi$ mesons in their decay to two muons. Both the $J/\psi$ mesons have been required to have rapidity and transverse momentum lying respectively in the ranges $2 < y < 4.5$ and $p_{T} < 10$ GeV/$c$.\\
The signal yield is determined fitting the invariant mass distribution of the first muon pair in bins of the second muon pair and correcting the number of signal events by the total efficiency. \\
The total efficiency is factorized in three different terms
\begin{equation}
\varepsilon _{J/\psi J/\psi}^{tot} = \varepsilon _{J/\psi J/\psi}^{sel \& reco \& acc} \times \varepsilon _{J/\psi J/\psi}^{\mu ID} \times \varepsilon _{J/\psi J/\psi}^{trg},
\end{equation}
where $\varepsilon _{J/\psi J/\psi}^{sel \& reco \& acc}$ is the acceptance, selection and reconstruction efficiency, $\varepsilon _{J/\psi J/\psi}^{\mu ID}$ is the efficiency of the muon identification and $\varepsilon _{J/\psi J/\psi}^{trg}$ is the trigger efficiency. To take into account the distortion due to the unknown $J/\psi$ polarization $\varepsilon _{J/\psi J/\psi}^{sel \& reco \& acc}$ is a function of the $J/\psi$ $\cos \theta$ where $\theta$ is the angle between the $\mu ^{+}$ in the $J/\psi$ center of mass frame and the Lorentz boost from the laboratory frame to the $J/\psi$ frame.\\
The corrected invariant mass distribution of the first muon pair in bins of the second muon pair is shown in Fig.\ref{2jpsiyield}, in three bins of $J/\psi$ transverse momentum in a particular bin of rapidity.\\
The double $J/\psi$ cross section is estimated to be
\begin{equation}
\sigma _{J/\psi J/\psi} = \frac{N^{corr} _{J/\psi J/\psi}}{\mathcal{L} \, \mathcal{B} ^{2}(J/\psi \rightarrow \mu ^{+} \mu ^{-})} = \left( 5.1 \pm 1.0 (stat.) \pm 1.1 (sys.) \right)  \mathrm{nb},
\end{equation} \label{2jpsires}
where $N^{corr} _{J/\psi J/\psi}$ is the efficiency corrected signal yield, $\mathcal{L} = 37$ pb$^{-1}$ is the integrated luminosity and $\mathcal{B} (J/\psi \rightarrow \mu ^{+} \mu ^{-})$ is the branching ratio of the $J/\psi$ decay into a muon pair.\\
The uncertainty are respectively statistical and systematic. The main contributions to the systematic uncertainty come from the tracking and trigger efficiency and from the unknown $J/\psi$ polarization.\\
The experimental result obtained by LHCb has been compared with the theoretical contribution calculated in color-singlet model, from the SPS and the DPS. The two contributions, estimated in the LHCb acceptance, are listed together with the related uncertainties, in Tab.\ref{tab:2jpsith} \cite{2jpsi}. The sum of the two contributions is in agreement with the experimental value reported in Eq.\ref{2jpsires}, although the uncertainties on the theoretical expectations are too large to draw a definite conclusion on the production mechanism.

\begin{table}[t]
\begin{center}
\begin{tabular}{l|cc}  
Model &  Cross section (nb) &  Uncertainty \\ \hline
 Single Parton S.  &   4.15     &     30\% \\
 Double Parton S. contribution &   2     &     50\% \\ \hline
\end{tabular}
\caption{SPS and DPS contribution to the double $J/\psi$ production cross section estimated in the LHCb acceptance. The theoretical uncertainties are also listed.}
\label{tab:2jpsith}
\end{center}
\end{table}

\begin{figure}
\begin{minipage}{.49\textwidth}
\centering
\includegraphics[width=1.\textwidth]{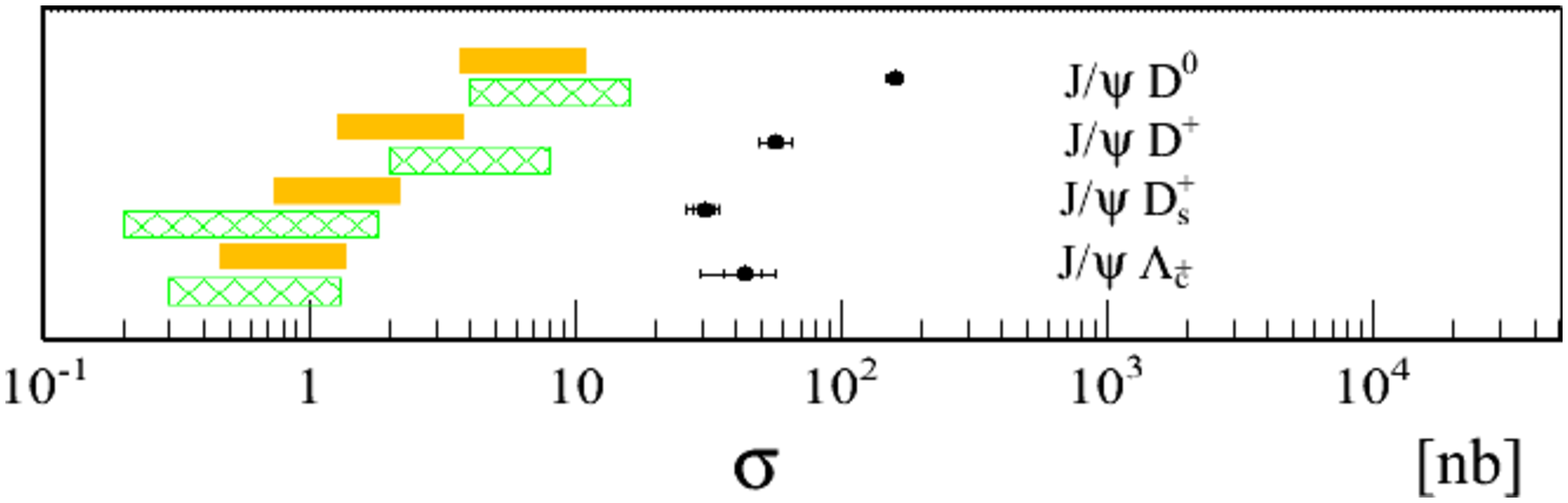}
\caption{$J/\psi$ production cross section with associated open charm. The experimental results (black points) are compared with the gluon-gluon fusion expectation (yellow and green areas \cite{SPS1, SPS2, SPS3}).}\label{jpsiD}
\end{minipage}
\begin{minipage}{.49\textwidth}
\centering
\includegraphics[width=1.\textwidth]{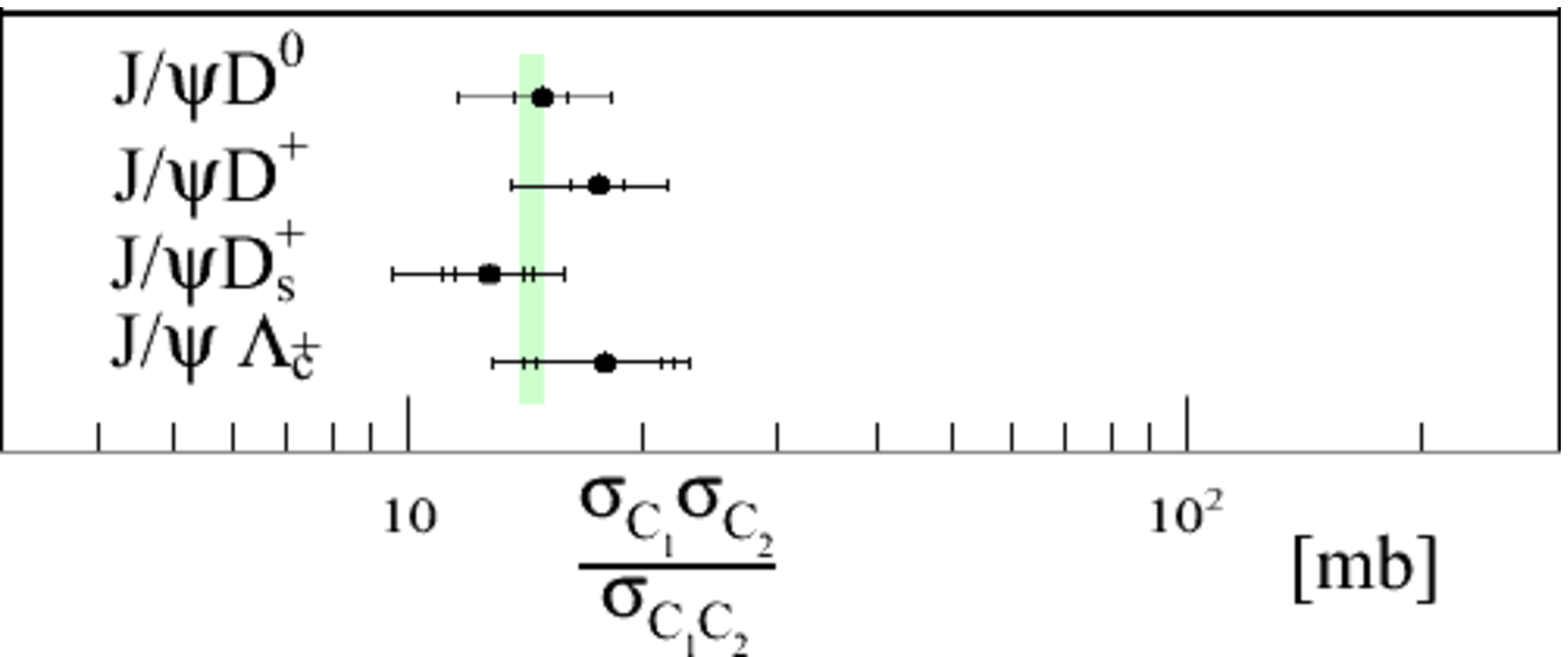}
\caption{Results for the prompt open charm cross sections and double open charm cross section ratio compared with the theoretical expectation computed with the DPS approach \cite{DPS1, DPS2, DPS3}.}
\label{jpsiDratio}
\end{minipage}
\end{figure}
The production cross sections of a $J/\psi$ meson associated with an open charm hadron, $D^{0}$, $D^{+}$, $D^{+}_{s}$ or $\Lambda _{c}^{+}$ have been measured using 355 pb$^{-1}$ out of the 2011 datasets. As control channels, the $c\overline{c}$ events with two open charm hadrons reconstructed in the LHCb acceptance have been also studied. The $J/\psi$, $D^{0}$, $D^{+}$, $D^{+}_{s}$ and $\Lambda _{c}^{+}$ hadrons have been reconstructed through the following decays: $J/\psi \rightarrow \mu ^{+} \mu ^{-}$, $D^{0} \rightarrow \pi ^{+} K ^{-}$, $D^{+} \rightarrow \pi ^{+} \pi ^{+} K ^{-}$, $D^{+}_{s} \rightarrow \pi ^{+} K ^{+} K ^{-}$, $\Lambda ^{+}_{c} \rightarrow p \pi ^{+} K ^{-}$.\\
In Fig.\ref{jpsiD} (taken from Ref.\cite{2jpsiopen}) the results for the production cross section of the $J/\psi D$ processes are shown, compared with the theoretical expectation estimated with gluon-gluon fusion model. In Fig.\ref{jpsiDratio} (from Ref.\cite{2jpsiopen}) the ratios of the product of the prompt open charm cross sections and the double open charm cross section show a good agreement with the theoretical expectation from the DPS, assuming the effective cross section measured in multi-jet events at Tevatron \cite{CDFopen}. In the plots the inner error bars represent the statistical uncertainties while the outer error bars are the sum in quadrature of the statistical and systematic uncertainties. The main contributions to the systematics are coming from the trigger efficiency, the unknown $J/\psi$ polarization and the luminosity.

\section{Conclusion}
LHCb has provided many contributions in understanding the quarkonium production mechanism in particular with the measurement of the $J/\psi$ and $\chi _{c}$ production. Also the double $J/\psi$ and $J/\psi$ associated with an open charm hadron production has been investigated, which provide a useful test for the SPS and DPS approach. The measurement of the polarization of the prompt $J/\psi$ component is ongoing and it will provide a critical test for the validity of the color-singlet and color-octet models.

\end{document}